\documentclass[twocolumn,pra,superscriptaddress]{revtex4}
\usepackage{amsmath}
\usepackage{graphicx}
\usepackage{amssymb}
\usepackage{dcolumn}
\usepackage{mathrsfs}
\usepackage{bm}
\usepackage{hyperref}
\hypersetup{
    colorlinks=true,
    linkcolor=blue,
    filecolor=blue,
    urlcolor=blue,
    citecolor=blue,
}
\makeatletter

\newcommand{\Rmnum}[1]{\expandafter\@slowromancap\romannumeral #1@}
\makeatother

\begin{document}

\title{Increase the Efficiency of Post-selection in Direct Measurement of Quantum Wave Function}
\author{Yong-Li Wen}
\affiliation{National Laboratory of Solid State Microstructures
and School of Physics, Nanjing University, Nanjing 210093, China}

\author{Shanchao Zhang}
\affiliation{Guangdong Provincial Key Laboratory of Quantum Engineering and Quantum Materials, School of Physics and Telecommunication Engineering, South China Normal University, Guangzhou 510006, China}

\affiliation{Guangdong-Hong Kong Joint Laboratory of Quantum Matter, Frontier Research Institute for Physics, South China Normal University, Guangzhou 510006,
China}

\author{Hui Yan}
\affiliation{Guangdong Provincial Key Laboratory of Quantum Engineering and Quantum Materials, School of Physics and Telecommunication Engineering, South China Normal University, Guangzhou 510006, China}

\affiliation{Guangdong-Hong Kong Joint Laboratory of Quantum Matter, Frontier Research Institute for Physics, South China Normal University, Guangzhou 510006,
China}

\author{Shi-Liang Zhu}
\email{slzhu@nju.edu.cn}
\affiliation{Guangdong Provincial Key Laboratory of Quantum Engineering and Quantum Materials, School of Physics and Telecommunication Engineering, South China Normal University, Guangzhou 510006, China}

\affiliation{Guangdong-Hong Kong Joint Laboratory of Quantum Matter, Frontier Research Institute for Physics, South China Normal University, Guangzhou 510006,
China}

\date{\today}

\begin{abstract}
Direct weak or strong measurement of quantum wave function has been demonstrated based on the post-selection; however, the efficiency of the measurement is greatly limited by the success probability of the post-selection. Here we propose a modified scheme to directly measure photon's wave function by simply inserting a liquid crystal plate after the coupling between the pointer and wave function. Our modified method can significantly increase the efficiency of the post selection.  Numerical simulations demonstrate that our proposal can have a greater efficiency in higher resolution and higher fidelity.

\end{abstract}

\maketitle

\section{Introduction}

A wave function is the fundamental concept  in quantum mechanics, and its determination is therefore of crucial importance in many applications.  However, in the traditional methods of quantum measurement developed by Von Neumann \cite{vonNeumann1955}, we can only obtain the modulus square of the wave function which determines the probabilities of all kinds of measurement outcomes. For a long time, quantum-state tomography has been considered as an operational measurement method to reconstruct the wave function \cite{Cramer2010,Resch2005,Smithey1993}.
 Quantum-state tomography is a conventional way to construct a quantum wave function, but it usually requires
plenty of projection measurements onto several bases sets, and thus it is also considered as an indirect measurement method with a limitation in efficiency and concision. Recently, an alternative operational definition of the wave function was proposed in Ref.\cite{Lundeen2011} based on the weak measurement \cite{Aharonov1988,Dressel2014,Ritchie1991} and post-selection. The method has been used to first measure the transverse wave function of photons, and then it has been extensively applied for the measurement of quantum state \cite{Salvail2013,Malik2014,Kocsis2011,Thekkadath2016,Lundeen2012,Bolduc2016,Qin2017,Bamber2014,Shojaee2018,Fischbach2012,Mirhosseini2014}. The method has also been generalized from the weak measurement to the strong measurement \cite{Vallone2016,Denkmayr2018,Calderaro2018,Pan2019,Zhang2020}.  An alternative direct measurement strategy based on a $\delta$-quench probe was developed: by quenching its complex probability amplitude one by one ($\delta$ quench) in the given basis, one can
directly obtain the quantum wave function of a pure ensemble by projecting the quenched state onto a
post-selection state \cite{SZhang2019}. However, the efficiency of these direct measurement methods is limited by the success probability of the post-selection, which is usually pretty low in a practical experiment, and thus increasing the efficiency of the post-selection is in demand.

In this work, we propose a modified method to directly measure photon's wave function by simply inserting a liquid crystal plate after the coupling between the pointer and wave function, and we show that our modified method can significantly increase the efficiency of the post selection.
We carefully analyze the measurement efficiency of the first experiment in which the transverse wave function of a photon has been measured with direct measurement method \cite{Lundeen2011}, and find that the efficiency is mainly dependent on the success probability of the
post-selection. In the
previous direct measurement scheme \cite{Lundeen2011,Vallone2016} , the photons
would be post-selected to a zero transverse momentum state, but the distribution in zero momentum is pretty low, especially under the conditions of high resolution and high measurement fidelity. By simply inserting a liquid crystal plate ( a Fourier
transformation lens which performs a Fourier transformation only on the polarization state $|1\rangle$), all signal part of the photons defined later can be zero momentum and thus become the post-selection state. Therefore, the post-selection would not decrease the measurement efficiency.  Numerical simulations demonstrate that our proposal can have a greater efficiency in higher resolution and higher fidelity. Our modified method can thus greatly increase the measurement efficiency of the direct measurement of the quantum wave function.

 This paper is organized as follows. In Section \ref{sec_model}, we review a general method of the direct measurement of wave function, including the previous weak and strong measurements as examples. The efficiency of the method is discussed. In section \ref{sec_our}, we propose a modified method to improve the efficiency of post-selection. Section \ref{sec_performances} investigates the performances of our proposed method. Finally, short discussion and  conclusion are given in Sec. \ref{sec_conclusion}.

\section{Direct measurement of wave function} \label{sec_model}

In this section, we present a general direct measurement method which includes the case under a weak coupling proposed in Ref.\cite{Lundeen2011} and strong coupling proposed in Ref. \cite{Vallone2016}, and then discuss the efficiency of this method.

We consider to measure a quantum wave function $|\psi\rangle=\sum_x\psi(x)|x\rangle$, where basis $\{|x\rangle\}$ are position eigenstates of a discretized segment.  A measurement can be seen as the coupling between an apparatus and a physical system that result in the translation of a pointer. Here we study the case of a photon spatial wave function, and the pointer can be represented by the polarization of the photons.  The coupling between an observable and the pointer can be written as a Hamiltionian $\hat{H}=g\hat{\pi}_{x}\hat{\sigma}_{y}$, where $\hat{\pi}_{x}=|x\rangle\langle x|$, $g$ is the coupling strength, and $\hat{\sigma}_{x,y,z}$ in this paper represent the Pauli matrices. The evolution operator with arbitrary coupling strength can be written as
\begin{equation}
\begin{aligned}
 \hat{U}(\theta)&=\exp[-ig\hat{\pi}_{x}\hat{\sigma}_{y}t/\hbar]\\
 &=\hat{I}_{\pi} \hat{I}_{\sigma}-2\sin^{2}\frac{\theta}{2}\hat{\pi}_{x} \hat{I}_{\sigma}-\sin \theta\left(i\hat{\pi}_{x} \hat{\sigma}_{y}\right),
 \end{aligned}
\end{equation}
where $\theta=gt/\hbar$,  $\hat{I}_{\pi}$ is the identity operator of position, and $\hat{I}_{\sigma}$ is the identity operator of the pointer. Suppose the initial state of the pointer is $|0\rangle$, the whole system after the coupling can then be written as
\begin{equation}\label{U}
\begin{aligned}
\hat{U}(\theta)|\psi\rangle|0\rangle=&[|\psi\rangle-2\sin^{2}\frac{\theta}{2} \psi(x)|x\rangle]|0\rangle-\sin \theta \psi(x)|x\rangle|1\rangle.
\end{aligned}
\end{equation}
If the system is projected to a momentum state $|p\rangle$, we have,
\begin{equation}
\label{fp}
\begin{aligned}
|f\rangle &=\langle p|\hat{U}(\theta)|\psi\rangle|0\rangle\\&=[\varphi_p-2\beta\sin^{2}\frac{\theta}{2}\psi(x)]|0\rangle-\beta\sin \theta \psi(x)|1\rangle,
\end{aligned}
\end{equation}
where $\varphi_p=\langle p|\psi\rangle$ and $\beta=\langle p|x\rangle$.
After the post-selection, the wave function can be obtained as \cite{Lundeen2005},
\begin{equation}\label{readout}
\begin{aligned}
&\frac{1}{2}\left[\left\langle f\left|\hat{\sigma}_{x}\right| f\right\rangle-i\left\langle f\left|\hat{\sigma}_{y}\right| f\right\rangle\right]-2\beta^{2}\sin^{2}\frac{\theta}{2}\sin \theta\langle f|\hat{P}_{1}|f \rangle.\\
=&\frac{1}{2}[(P_{+}-P_{-})-i(P_{L}-P_{R})]-2\beta^{2}\sin^{2}\frac{\theta}{2}\sin \theta P_{1}\\
=&-{\beta\sin \theta} \varphi^{*}_p\psi\left(x\right).
\end{aligned}
\end{equation}
where $P_{+}=\langle f|+\rangle\langle +|f\rangle$, $P_{-}=\langle f|-\rangle\langle -|f\rangle$, $P_{L}=\langle f|L\rangle\langle L|f\rangle$, $P_{R}=\langle f|R\rangle\langle R|f\rangle$, and $P_{1}=\langle f|1\rangle\langle 1|f\rangle$ are the probabilities of  $|f\rangle$ projecting on the following pointer states,
\begin{equation}\label{states}
\begin{aligned}
&|+\rangle=\frac{1}{\sqrt{2}}(|1\rangle+|0\rangle),\quad |-\rangle=\frac{1}{\sqrt{2}}(|1\rangle-|0\rangle),\\
&|L\rangle=\frac{1}{\sqrt{2}}(|1\rangle+i|0\rangle),\quad |R\rangle=\frac{1}{\sqrt{2}}(|1\rangle-i|0\rangle).\\
\end{aligned}
\end{equation}
If the coupling strength is sufficiently weak to neglect the second order of $\theta$, equation (\ref{readout}) will lead to the result of weak measurement studied in Ref.\cite{Lundeen2011},
\begin{equation}\label{weak}
\begin{aligned}
\frac{1}{2}[(P_{+}-P_{-})-i(P_{L}-P_{R})]=-{\beta\sin \theta} \varphi^{*}_p\psi\left(x\right).
\end{aligned}
\end{equation}
However, if $\theta$ is too small, the intensity of the final readout of the measurement may be very low. When $\theta=\pi/2$, it becomes the direct strong measurement case investigated in Ref.\cite{Vallone2016},
\begin{equation}\label{strong}
\begin{aligned}
&\frac{1}{2}\left[(P_{+}-P_{-})-i(P_{L}-P_{R})\right]-|\beta|^{2}P_{1}
=-{\beta} \varphi^{*}_p\psi\left(x\right).
\end{aligned}
\end{equation}
Benefit from the higher coupling strength, this direct strong measurement  method  has a better precision in wave function measurement. We will focus on the strong  coupling case in this paper.

The prefactors ${\beta} \varphi^{*}_p$ of the wave function $\psi (x)$ in  Eq. (\ref{strong}) determine the efficiency of the direct measurement.
 Both $\varphi^{*}_p$ and $\beta$ depend on the success probability of the post-selection. In the experiment reported in Ref.\cite{Lundeen2011}, $\varphi^{*}_p$ is the probability amplitude at zero momentum of the unknown state $|\psi\rangle$, and  it is an $x$ independent constant. The $\beta$ is a value related to the width of both the measuring step of $x$ and the post-selection. The smaller of the value of the $\beta$, the lower of the success probability of the post-selection. In many situation, $\beta$ is the major loss of the success probability of the post-selection. Therefore, optimizing the value of $\beta$ plays a key role in improving the efficiency of a direct measurement experiment.

\section{Proposal to improve the probability of post-selection} \label{sec_our}

\begin{figure}[ptb]
\includegraphics[width=8cm]{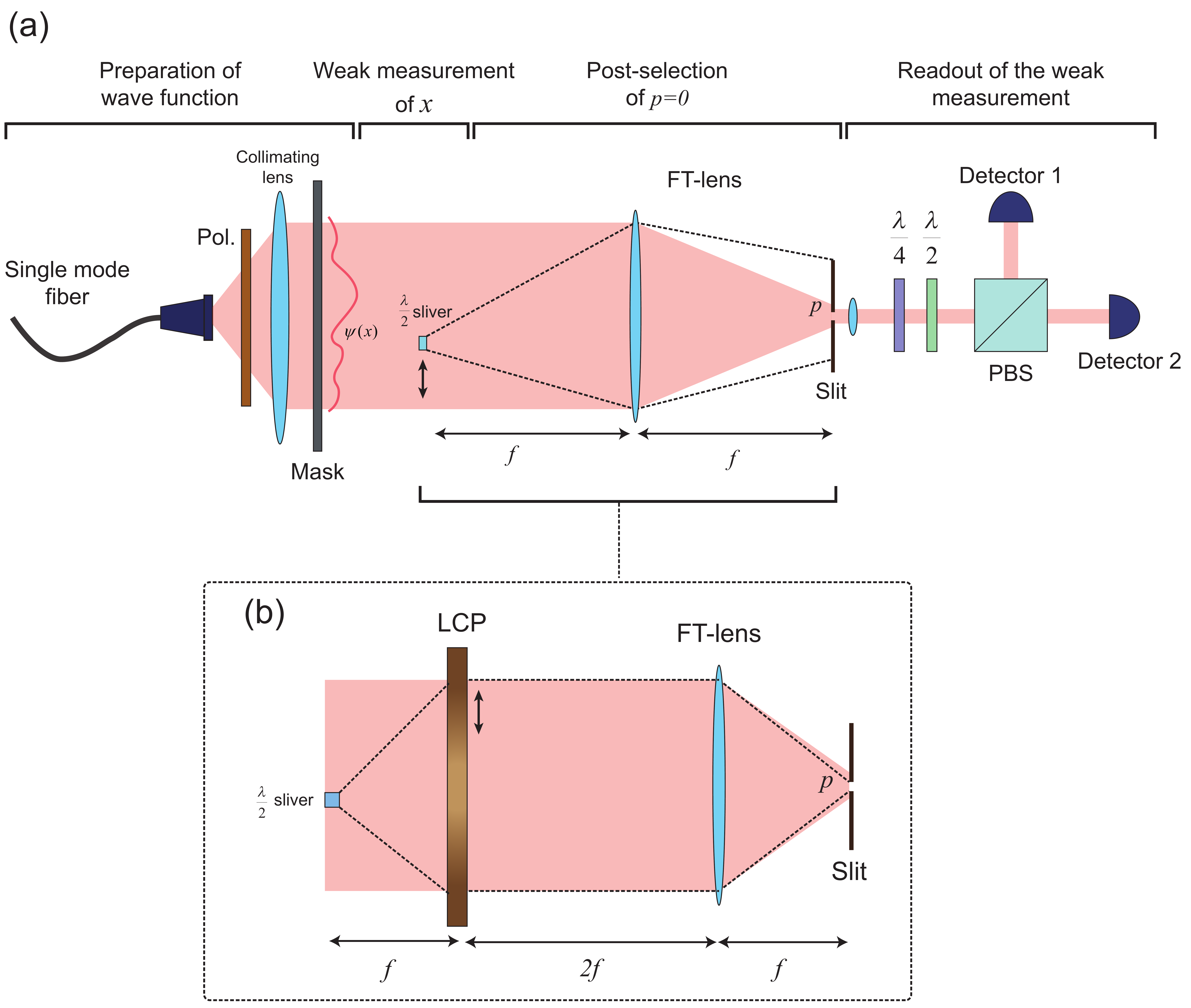}
\caption{{(a)} Schematic setup of the experiment  proposed in Ref.\cite{Lundeen2011} to directly measure quantum wave function. The experiment has four sequential steps: state preparation, interaction of the probe system and the pointer, post-selection of the system, and readout the measurement results.
{(b)} Our modified method to increase the efficiency of the post-seletion. 
}
\end{figure}

As a concrete example, we address our method to optimize the efficiency of the post-selection based on a previous optical experiment \cite{Lundeen2011}, which is schematically	
 shown in Fig.1. In the experiment, the researchers considered a photon
travelling along the $z$ direction, and the aim is to directly measure the $x$ wave function  of the photon.

An ensemble of single photons  are emitted from a
single mode fibre and collimated. The polarization of the photons is chosen as the pointer in the measurement. The polarizer (Pol.) is used to initialize the polarization state of the photons to be $|0\rangle$.  The photons pass through a collimating lens to make the transverse mode of the photons a nealy parallel propagating Gaussian beam. A phase mask placing after the collimating lens generates the initial quantum wave function $\psi(x)$. A half wave plate ($\lambda/2$) sliver placing at the transverse position $x$ performs the coupling between the pointer and the wave function $\psi(x)$. We can consider that the photons have been divided into two parts: the unperturbed part in the pointer state $|0\rangle$, and the signal part in the pointer state $|1\rangle$. As Eq.(\ref{fp}) shows, the signal part is carrying the information of wave function $\psi(x)$. In the previous direct measurement scheme\cite{Lundeen2011,Vallone2016}, the photons would be post-selected to a transverse momentum state of $|p=0\rangle$ after the coupling. The dash line in Fig.1a shows the propagation of the signal part. The photons modulated by the $\lambda/2$ sliver evolve approximately like a  point-source light. Then the Fourier transformation lens (FT-lens) transforms the light to a momentum spectrum at the focal plane. As the measuring region $\delta x$ is small, the uncertainty of the transverse momentum of the signal part is very large comparing with the post-selected momentum $p \approx 0$. It leads to a consequence that only a small portion of the signal part can be successfully selected by the slit. Thus it largely limits the efficiency of the measurement.

After the slit, a lens is used to collimate the outcoming light. Then the quarter wave plate (QWP), half wave plate (HWP) and the PBS  are used to project the final pointer state to four diagonal bases respectively. This step is to read out the expectation values  $P_{\pm}$, $P_{L,R}$, and $P_1$ in Eq.(\ref{strong}).

 We here propose a modified method to increase the efficiency of the post-selection.
    Our modification of the experimental proposal is shown in Fig.1b: a liquid crystal plate (LCP) is inserted between the $\lambda/2$ sliver and the FT-lens. In our modified scheme, the photons first transmit the LCP after the coupling of the half wave plate sliver. This LCP can be considered as a fourier transformation lens which can perform a Fourier transformation only on the polarization eigen state $|1\rangle$ (\ref{lcpapp}).
     If we adjust the longitudinal distance between the $\lambda/2$ sliver and the LCP to be $f$ which is the focal length of this LCP, the light of the signal part will be transformed to be a plane wave. This plane wave is an eigenstate of momentum, $|p\rangle=e^{ipx/\hbar}$. If the transverse position of the LCP is fixed, the position $x$ of the signal part modulated by the $\lambda/2$ sliver will be transformed to different eigen value of $p$. For the purpose of making this eigen value of momentum to be zero at different measuring point in $x$, we can adjust the transverse position of the LCP accompanied by the change of measuring position on the sliver. Then the measuring position $x$ can be always at the same transverse position of the LCP. After this adjustment, the eigen value of the transformed wave plane can be constantly zero, $p_0=0$. At the longitudinal distance of $2f$ after the LCP, we place a normal FT-lens with a focal length of $f$. This FT-lens performs a Fourier transformation on both  polarization eigen states $|0\rangle$ and $|1\rangle$. Then the system can be expressed as,
\begin{equation}
\label{cprime}
|c'\rangle=[|\psi\rangle-\psi(x)|x\rangle]|0\rangle-\psi(x)|p_{0}\rangle|1\rangle
\end{equation}
At the focal plane of the FT-lens, the transverse distrubution of the light is relative to the momentum wave function of $|c'\rangle$. In Fig.1b, we can notice that the combination of the LCP and the FT-lens constitute a $4f$ imaging system. This $4f$-system transfers the signal part $\psi(x)|1\rangle$ from the measuring position to the focal plane of the FT-lens. Remarkably, the momentum of all signal part become $p_0=0$, so the $|p_0\rangle$ state  appears in Eq.(\ref{cprime}), in contrast to any moment $p$ in Eq. (\ref{fp}).

We use a slit placing at the center of the focal plane to project $|c'\rangle$ into a post-selected state of zero momentum $|p_{0}\rangle$. Then the final state of the pointer is given by,
\begin{equation}
\begin{aligned}
|f\rangle=&\langle p_{0}|c'\rangle\\
=&[|\varphi(0)-\beta\psi(x)|0\rangle-\psi(x)|1\rangle.
\end{aligned}
\end{equation}
 Following the calculation in the previous direct strong measurement scheme, the probe wave function can be obtained as,
\begin{equation}\label{e04}
\begin{aligned}
&\frac{1}{2}\left[\left\langle f\left|\hat{\sigma}_{x}\right| f\right\rangle-i\left\langle f\left|\hat{\sigma}_{y}\right| f\right\rangle\right]-|\beta|\langle f|\hat{P}_{1}|f \rangle
=-\varphi^{*}\left(0\right)\psi\left(x\right).
\end{aligned}
\end{equation}
 Comparing Eq. (\ref{strong}) and Eq. (\ref{e04}), we  notice that the coefficient $\beta$ has been eliminated in our modified scheme. In other words, our measurement scheme has a magnification $\mathcal{M}=1/|\beta|$ in the final read out of the experiment.
\begin{figure}[ptb]
\includegraphics[width=6cm]{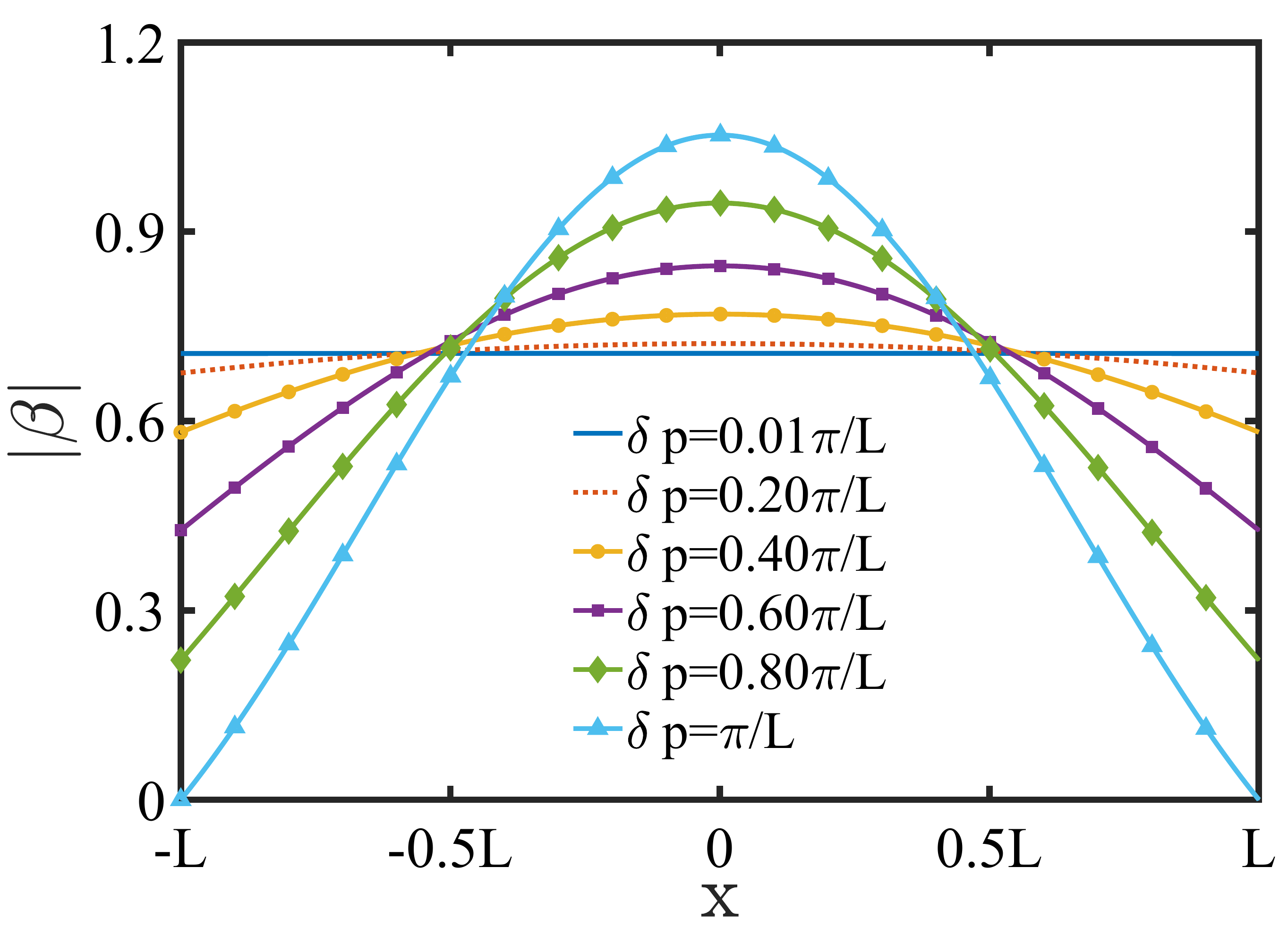}
\caption{\label{fig:sys} \textbf{(a)}. The factor $|\beta|$ as a function of $x$. $\delta p$ can be chosen in $[0,\pi/L]$, where $[-L,L]$ is the measuring range of $\psi(x)$ in $x$.  The function $|\beta|^{2}$ of each curves has been normalized in the measuring range.
}
\end{figure}

\begin{figure}[ptb]
\includegraphics[width=6cm]{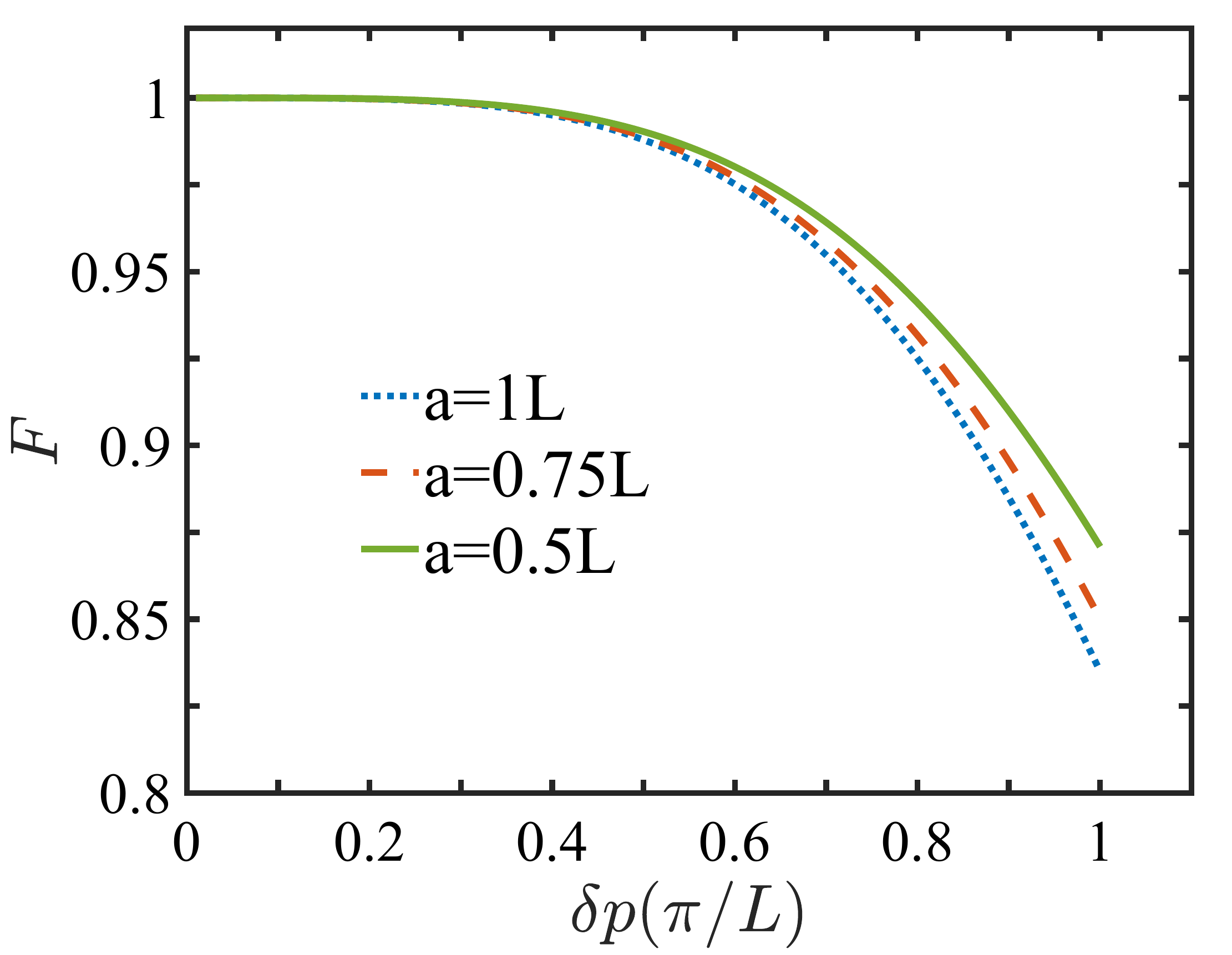}
\caption{\label{fig:sys}  The fidelity F of the measured wave function as a function of $\delta p$. The waists of the Gaussian wave packets $a=1.00 L, 0.75 L, 0.50 L$.  }
\end{figure}

\section{Performances of the improved method}\label{sec_performances}

 We now turn to analyse the key magnification factor $\mathcal{M}=1/|\beta|$ with $\beta =\langle p|x\rangle$. For convenience, we usually choose a post-selected state to make $\beta$ a constant. A state meet this condition is mutually unbiased bases. In a discrete system, two bases $\{b\}$ and $\{a\}$ are mutually unbiased \cite{Durt2010} if the square of the magnitude of the inner product between any basis states equals the inverse of the dimension $d$, $|\langle b|a\rangle|^{2}=1/d$. For a typical mutually unbiased bases in an infinite dimension Hilbert space, position state $|x\rangle$ and momentum state $|p\rangle$, the inner product is in a plane wave formulation, $\langle p|x\rangle=\frac{1}{\sqrt{2\pi \hbar}}e^{ipx/\hbar}$. In previous works of direct measuring the position wave function, $\{|x\rangle\}$ are recast to $d$-dimension position eigenstates. Then the position quantum state is written in a discrete formulation, $|\psi\rangle=\sum_{x=1}^{d} \psi_{x}|x\rangle$. After this discretization, the post-selected state of zero momentum can be expressed as a mutually unbiased bases, $\left|p_{0}\right\rangle=(1 / \sqrt{d}) \sum_{x=1}^{d}|x\rangle$, so we can obtain $\beta=1/\sqrt{d}$. Therefore, the magnification of our modified proposal is $\mathcal{M}=\sqrt{d}$. Since $d$ is the number of the measuring points, the magnification $\mathcal{M}$ is higher in high resolution of measurement.

 In a practical experiment,  $|p_{0}\rangle$ is not an exact zero momentum state, it should be an integral of momentum eigen states in the range  $[-\delta p,\delta p]$, where $\delta p$ is related to the width of the slit. Under this condition, $\beta$ can be recast as $\beta=\int^{\delta p}_{-\delta p}\frac{1}{\sqrt{2\pi \hbar}}e^{ipx/\hbar}dp$. In Fig.2, we show $\beta$ as a function of $x$ for different values of $\delta p$. It is shown that $|\beta|$ has a good flatness in the range $x\in[-L,L]$  when $\delta p$ is much smaller than $\pi/L$. So to  make $\beta$ a nearly $x$ independent value, $\delta p$ has to be sufficiently small. Simultaneously, the success probability of post-selection would decrease as $\delta p$ decreases.   Therefore, $\delta p$ effects the precision of measuring the wave function.


There is a trade off between high efficiency of post-selection and the precision of the measurement. We thus further evaluate how the value $\delta p$ effects the accuracy of the measurement. We denote the ideal wave function in the preparation as $\psi_{G}(x)$, which is a Gaussian wave packet $\psi_{G}(x)=[\frac{a}{a+it/m}]^{\frac{3}{2}}\exp{[-\frac{x^{2}}{2(a+it/m)}]}$. Here $a$ is the waist of the wave packet, $t$ is the propagation time from the preparation to the moment of the coupling on $\lambda/2$ sliver, $m=\frac{2\pi \hbar}{\lambda c}$ with $c$ being the velocity of light and $\lambda$ being the wave length of the light is the effective mass of the light.  If we only consider the errors caused by the width of the post-selection $\delta p$, the measured wave function can then be written as $\psi_{m}(x)=\beta(x)\psi_{G}(x)$. The fidelity of these two wave function can be used to describe the accuracy of the measurement, i.e.,
\begin{equation}\label{fidelity}
\begin{aligned}
F=|\langle \psi_{m}|\psi_{G}\rangle|^{2}.
\end{aligned}
\end{equation}
 We calculate the fidelity $F$ as a function of $\delta p$  for various  waist $a$, and the results are shown in Fig.3. The range of $\delta p$ is $[0,\pi/L]$, and the measuring position $x \in [-L,L]$. These two ranges satisfy the condition of $2\pi/(2L)=\pi/L$, which  is required for the discrete Fourier transformation.  It is clearly that the fidelity decreases with increasing $\delta p$.

On the other hand, the width of the measuring step $\delta x$ also effects $\beta$.  $\delta x$ can be considered as the uncertainty of transverse position. According to Heisenberg's uncertainty principle, the uncertainty of transverse momentum becomes larger when $\delta x$ decreases. For a given $\delta p$, higher resolution of the measurement will lead to a lower efficiency of post-selection.

\begin{figure}[ptb]
\includegraphics[width=6cm]{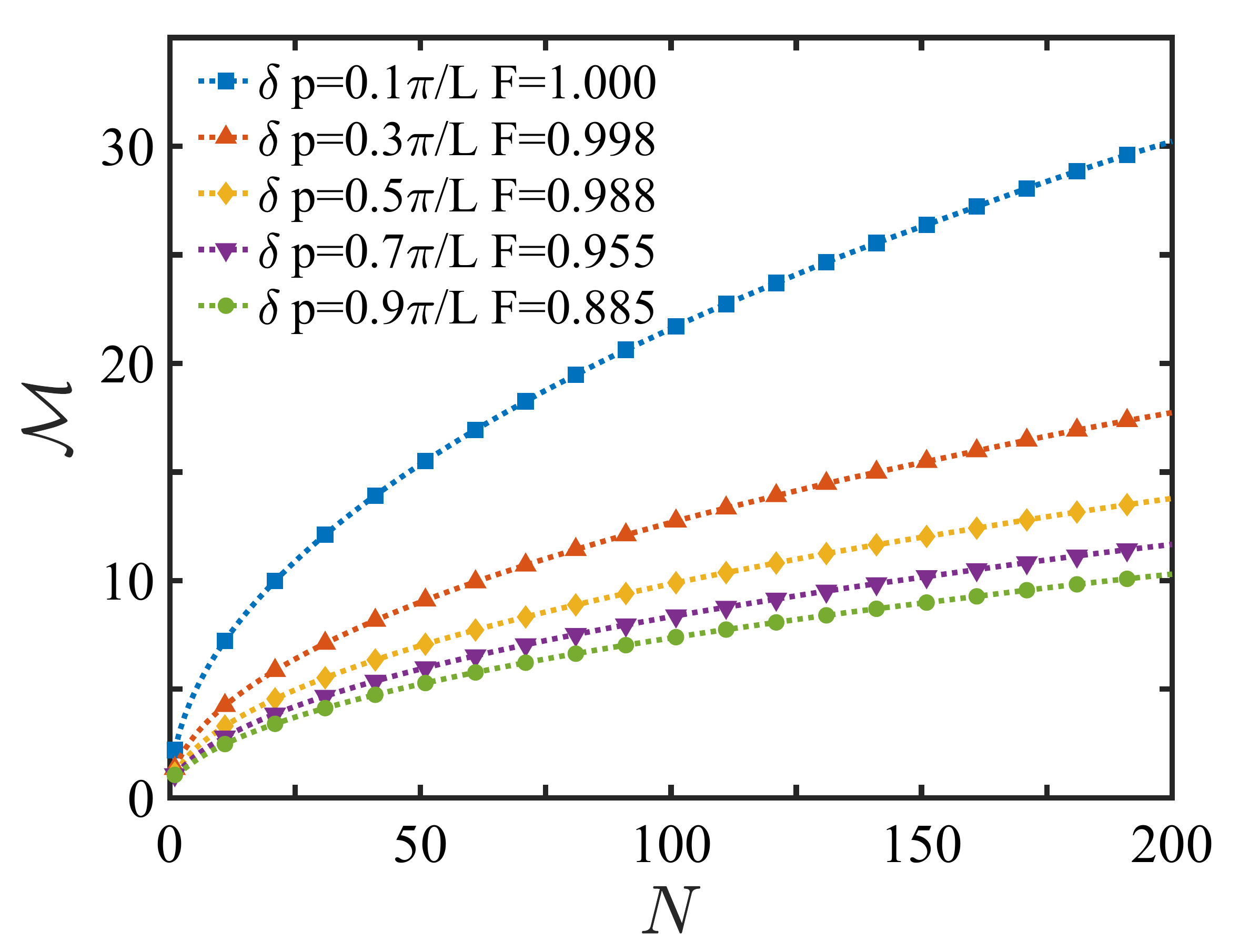}
\caption{\label{fig:sys} \textbf{(a)}. The magnification factor $\mathcal{M}$ as a function of the measuring point $N$.  The waist of the Gaussian wave packet $a=L$. }
\end{figure}

For the fixed $\delta x$ and $\delta p$, $\beta$ can be written as a double integral formulation $\beta=\int^{\delta x}_{-\delta x}dx\int^{\delta p}_{-\delta p}\frac{1}{\sqrt{2\pi \hbar}}e^{ipx/\hbar}dp$. Fig. 4 shows the magnification factor $\mathcal{M}=1/|\beta|$ as a function of number $N$ which is the measuring points with a given measuring range $x\in[-L,L]$. The magnification factor is higher when the number $N$ increases. Furthermore, our proposed method has a higher efficiency for higher resolution and fidelity.

\section{Discussion and Conclusion} \label{sec_conclusion}

  We discuss the improvement of the efficiency under the conditions in Ref.\cite{Lundeen2011}, where the parameters  $L=30mm$, $\delta x=1mm$, $f=1000mm$, $l_{slit}=15\mu m$, and $\lambda=800nm$. We can estimate the width of measuring step $\delta x\approx0.033L$ and the post-selection $\delta p\approx0.563(\pi/L)$ by using the formula $\delta p=\pi l_{slit}/f\lambda$. With these two parameters ($\delta x$ and $\delta p$), we can obtain $|\beta|\approx0.138$ by numerical calculation. Therefore, if we modify the experimental setup in Ref.\cite{Lundeen2011} with our proposal, it would have magnification of $\mathcal{M}=1/|\beta|\approx 7.24$ in efficiency increasing.

  In summary, we have proposed a modified scheme to directly measure quantum wave function by simply adding a LCP after the coupling between the pointer and the wave function. Our modified method can significantly increase the efficiency of the post selection.  Numerical simulations have shown that our proposal can have a greater efficiency in higher resolution and fidelity.

\acknowledgments

This work was supported by  the Key-Area Research and Development Program
of GuangDong Province (Grant No. 2019B030330001),  the National Natural Science Foundation of China (Grants No. 12074180, No. U1801661), the Key Project of Science and Technology of Guangzhou (Grants No. 201804020055 and No. 2019050001), and the National Key Research
and Development Program of China (Grant No. 2016YFA0301800).

\begin{appendix}
\section{PHASE MODULATION OF the LCP}\label{lcpapp}
In the theory of Fourier optics, the Fourier transformation lens (FT-lens) performs a phase retardation on a wave front. In dimension $x$, the phase transformation function is given by,
\begin{equation}\label{psi}
\begin{aligned}
U_{ft}=e^{-i\frac{\pi x^{2}}{\lambda f}},
\end{aligned}
\end{equation}
where $\lambda$ is the wave length of the light and $f$ is the focal length of the FT-lens. The LCP can be considered as a FT-lens only act on the polarization state $|1\rangle$. Now we return to analyze our experimental proposal. The signal part in Eq. (\ref{fp}) with $\theta=\pi/2$ is $|\psi_{s}\rangle=\psi(x)|x\rangle$. Besides, the free evolution Hamiltonian of photon is $\hat{H}=\frac{\hat{p}^{2}}{2m}$, where $m=\frac{2\pi\hbar}{\lambda c}$ is the equivalent mass of the photon. Then the signal part evolve for a distance of $f=tc$, and the wave function after this evolution is given by,
\begin{equation}\label{psi}
\begin{aligned}
\psi'(x')&=\langle x'|e^{-i\frac{p^{2}}{2m}\frac{t}{\hbar}}\psi(x)|x\rangle\\&=\mathcal{A}\psi(x)e^{i\frac{mc(x'-x)^{2}}{2\hbar f}},
\end{aligned}
\end{equation}
where $\mathcal{A}$ is the normalization factor and $x'$ is the transverse position at $f$ away from the $\lambda/2$ sliver. At this moment, if the center of the LCP is at the position of $x$, it will performs a phase retardation of $e^{-i\frac{\pi (x'-x)^{2}}{\lambda f}}$ at $x'$. The wave function after the modulation of LCP is given by,
\begin{equation}\label{lcp}
\begin{aligned}
\psi_{L}(x')&=e^{-i\frac{\pi (x'-x)^{2}}{\lambda f}}\psi'(x')\\
&=\mathcal{A}\psi(x)e^{-i\frac{\pi (x'-x)^{2}}{\lambda f}}e^{i\frac{mc(x'-x)^{2}}{2\hbar f}}\\
&=\mathcal{A}\psi(x).
\end{aligned}
\end{equation}
We can notice that, $\psi_{L}(x')$ is a constant in $x'$. At this moment, the state in $x'$ can be written as $|\psi_{L}\rangle=\psi(x)\sum_{x'}\mathcal{A}|x'\rangle$. If we recast Eq. (\ref{lcp}) into momentum representation, it becomes the eigenstate with zero momentum $|\psi_{L}\rangle=\psi(x)|p=0\rangle$. With the modulation of the LCP, the eigenstate of transverse position $x$ is transformed to a zero transverse momentum state which is also the post-selection state in the measurement. Notably, state $|\psi_{L}\rangle=\psi(x)|p=0\rangle$ still carrying the wave function $\psi(x)$ as a coefficient.
\end{appendix}

\end{document}